\documentclass[12pt,epsf]{article}
\usepackage{cite}
\newlength{\abstractwidth}
\usepackage[dvips]{graphicx}
\usepackage{subfigure}
\usepackage{amssymb}

\renewcommand{\thefootnote}{\fnsymbol{footnote}}
\renewcommand{\thanks}[1]{\footnote{#1}} 
 \newcommand{\starttext}{
\setcounter{footnote}{0}
\renewcommand{\thefootnote}{\arabic{footnote}}}
\renewcommand{\theequation}{\thesection.\arabic{equation}}
\newcommand{\be}{\begin{equation}}
\newcommand{\bea}{\begin{eqnarray}}
\newcommand{\eea}{\end{eqnarray}}
\newcommand{\beq}{\begin{equation}}
\newcommand{\ee}{\end{equation}}
\newcommand{\eeq}{\end{equation}}

\newcommand{\<}{\langle}

\newcommand{\comment}[1]{}
\renewcommand{\>}{\rangle}
\def\ba{\begin{eqnarray}}
\def\ea{\end{eqnarray}}

\def\t0{$t=0$}
\def\14{{1\over4}}
\def\12{{1 \over 2}}

\def\d{\partial}

\def\h3{h^{3\over 2}}

\def\>{\rangle}
\def\<{\langle}

\def\0cc{$\Lambda = 0$}

\def\fa{{\ensuremath{f_a}}}
\def\d{{\ensuremath{\zeta}}}
\def\tl{{\ensuremath{t_\Lambda}}}

\begin{document}
\renewcommand{\theequation}{\thesection.\arabic{equation}}
\begin{titlepage}
\bigskip

\bigskip\bigskip\bigskip\bigskip

\centerline{\Large \bf {Anthropic Explanation of the Dark Matter Abundance}}

\bigskip\bigskip
\bigskip\bigskip

\centerline{\bf Ben Freivogel}
\medskip
\centerline{\small Department of Physics and Center for Theoretical
Physics} 
\centerline{\small 
University of California, Berkeley, California 94720, U.S.A.}
\medskip
\centerline{\small Lawrence Berkeley National Laboratory, 
Berkeley, California 94720, U.S.A.}
\medskip
\centerline{\small freivogel@berkeley.edu}

 \begin{abstract} I use Bousso's causal diamond measure to make a
 statistical prediction for the dark matter abundance, assuming an
 axion with a large decay constant $f_a \gg 10^{12}$ GeV.  Using a
 crude approximation for observer formation, the prediction agrees
 well with observation: 30\% of observers form in regions with less
 dark matter than we observe, while 70\% of observers form in regions
 with more dark matter.  Large values of the dark matter ratio are
 disfavored by an elementary effect: increasing the amount of dark
 matter while holding fixed the baryon to photon ratio decreases the
 number of baryons inside one horizon volume.  Thus the prediction is
 rather insensitive to assumptions about observer formation in
 universes with much more dark matter than our own.  The key
 assumption is that the number of observers {\it per baryon} is
 roughly independent of the dark matter ratio for ratios near the
 observed value.  \end{abstract}

\end{titlepage}
\starttext \baselineskip=18pt \setcounter{footnote}{0}

 

\setcounter{equation}{0}

\section{Introduction}
Our current understanding of string theory suggests a vast landscape, 
with eternal inflation leading to an infinite number
of pocket universes containing different low energy physics \cite{bp,
  kklt, lenny}. In this
context, a prediction requires a combination of landscape
statistics, the dynamics of eternal
inflation, and anthropic considerations. Some quantities may still
have a conventionally natural explanation, while
for others anthropic arguments are crucial. The most famous
example of a quantity whose observed value is well explained by
 anthropics
is the cosmological constant \cite{cc}.

It is natural to wonder whether the dark matter abundance has an
anthropic explanation. For example, models with a large axion decay
constant, $\fa \gg 10^{12}$ GeV, are attractive from the particle
physics point of view \cite{pq} and arise naturally in string
theory. However, in these theories the dark matter abundance is
naturally much larger than what we observe \cite{axbound}. Many
authors\cite{axanth, simeon,
  tarw, htw}\footnote{I apologize
  for missing references throughout this note. Please let me know if your work should
  be cited.}, beginning with Linde, 
have examined the question of whether the dark matter
abundance is typical
once anthropic weighting is taken into account. These authors have
reached a variety of conclusions. Due to the
difficulty in simulating universes different from our own and our
ignorance about the conditions necessary for life, it is
unclear whether a dark matter abundance of even 100 times the observed value
prevents observers from forming. Furthermore, even the most stringent
assumptions about the requirements for life \cite{tarw} lead to the
conclusion that the observed dark matter abundance is unusually small.

It is not possible to make a prediction in the landscape without
regulating the infinities. Eternal inflation produces an infinite
number of pocket universes of every type, and each pocket universe is
spatially infinite. One attractive recipe for regulating the
infinities 
 is the causal diamond measure of Bousso \cite{cd}, which explains the
 observed cosmological constant well \cite{harnik, by}. 

Here I focus on the simplest possible scenario: I fix all
parameters to their observed values and fix the axion decay constant
to a large value, $f_a \gg 10^{12}$ GeV. I assume that axions make up
all of the dark matter. The only parameter which is allowed to vary is
the initial axion misalignment angle $\theta_i$. In terms of
observable quantities, this means that the baryon to photon ratio is held
fixed, while the ratio of dark matter to ordinary matter is allowed to
vary.
In this simple context, the causal diamond measure just tells us to
count the number of observations within the backward lightcone of one
geodesic. In other words, we can get a finite answer from spatially
infinite universes by restricting our attention to one horizon
volume. (The ``horizon'' throughout this paper refers to everything within
the backward lightcone of one geodesic, and never to the apparent horizon.)

 In this note, I show that dark matter abundances much larger than
 what we observe are disfavored in the causal diamond measure by an
 elementary effect: increasing the amount of dark matter while holding
 fixed the baryon to photon ratio decreases the
 number of baryons inside our causal patch. Quantitatively, if \d\ is
 the ratio of dark matter to ordinary matter, the number of baryons
 $N_b$ in the causal patch is \be N_b \propto {1 \over 1 + \d} ~.  \ee

This may be surprising, because it seems that we are not subtracting
any baryons when we increase the amount of dark matter. To see that
the above formula is correct
at least at one time, consider the time \tl\ when the
cosmological constant begins to dominate the energy density. The
horizon volume at this time is set by the cosmological constant,
independent of the dark matter ratio \d. By definition, at the time
\tl\ the total matter density is equal to the vacuum energy density,
$\rho_\Lambda (\tl) = \rho_m (\tl)$, so the total amount of matter
inside the horizon at the time \tl\ is independent of \d. Therefore,
the number of baryons inside the horizon is $N_b (\tl) \propto 1/(1 +
\d)$. Incidentally, the time \tl\ has only a tiny dependence on \d.

 A similar effect is likely to be present in measures other than the
 causal diamond measure. For example, the scale factor measure
 \cite{dgsv} in the homogeneous approximation effectively weights
 vacua by the physical density of observers \cite{bfy}. The density of
 baryons at fixed time is proportional to $1/(1 + \d)$, so the same
 weighting factor is recovered.

While this effect is independent of the details of the dark matter, 
the prediction for the dark matter abundance depends on the 
prior distribution for the dark matter
density.
The amount of dark matter is determined by the initial axion
misalignment angle
 $\theta_i$, and symmetry guarantees that $\theta_i$ is randomly
distributed. 
For ``unnaturally'' small dark matter ratios, $\theta_i$ is near its
minimum, and $\d \propto \theta_i^2$. This leads to a 
prior distribution for the ratio of dark matter to ordinary
matter
\be
{d P \over d \d}  \propto {1 \over \sqrt{\d}}~\ \ {\rm (prior)}.
\ee

In principle, after computing the prior probability distribution  for
\d\ and the number of
baryons as a function of \d, one should compute the number of observations per baryon
as a function of \d. Here I follow Tegmark,
Aguirre, Rees, and Wilczek \cite{tarw} and assume that the number
of observations per baryon is roughly constant for dark matter ratios
in the range $2.5 < \d < 100$, while using the
approximation that outside of this range there are almost no observations per
baryon.  I find that 30\% of observers see less dark matter than we
do, while 70\% see more. Therefore, our observations are quite typical.
The
distribution is rather broad: in our approximation, 68\% of observers
see a dark matter ratio $\d \lesssim 15$, while 95\% of observers see
$\d \lesssim 65$

The results are rather robust, which is good because the conclusions
of \cite{tarw} are controversial. For example, Hellerman and Walcher
\cite{simeon} conclude that, to the best of our current understanding,
 observers may form up to $\d \approx
10^5$. I show that the statistical prediction here is only mildly
sensitive to assumptions about observer formation at large $\d$,
unlike in the conventional anthropic analysis: assuming that observers
form efficiently all the way up to $\d = \infty$ only has the effect
of changing the number of observers who see less dark matter
than we do from 30\% to 25\%.

While the prediction for the dark matter abundance agrees well with
observation, it may not be absolutely persuasive to antianthropicists
because the probability
distribution is rather broad and the dark matter ratio has already
been measured. Confirmation of these ideas could come from
observation, because axionic dark matter has distinctive observational
signals in the form of isocurvature perturbations \cite{htw, kapnel}.

In the next section, I review the prior distribution for the dark
matter abundance in high-scale axion models. In section 3, I compute
the number of baryons inside the horizon as a function of \d\ and
describe the approximation for the number of observers per baryon. In
section 4, I combine these results to get a statistical prediction
for the dark matter abundance. In section 5, I mention several
interesting future directions.

\section{Prior Distribution}
We focus on a small part of the landscape. We fix the axion decay
constant to a large value, $f_a \gg 10^{12}$ GeV, and assume that the
energy scale of inflation is significantly smaller than $f_a$. We 
consider the set of vacua exactly like
ours, but with varying initial axion misalignment angle $\theta_i$. To
be concrete, we could imagine that slow roll inflation begins by a
tunneling event from a metastable false vacuum, 
and that the shift symmetry of the axion is
unbroken in the entire regime of interest. In other words, the axion
is basically a spectator in the tunneling event. Then every time a
bubble forms the axion is roughly homogeneous throughout the interior
of the bubble, with a random initial misalignment angle.
 
Once the universe cools sufficiently, the potential for the axion
becomes important. The axion field is approximately homogeneous in one
horizon volume and acts like dark matter \cite{axbound}. The amount of dark matter
relative to baryonic matter is determined by the misalignment angle.
In models with a large axion decay constant, the
natural value for the dark matter abundance is much larger than what
we observe, so values near the observed value correspond to the axion
being rather close to the minimum of its potential. In this regime,
the potential is approximately quadratic, so the dark matter abundance
is
\begin{equation}
\d \propto \theta_i^2~.
\end{equation}
Therefore, the prior probability distribution for $\sqrt{\d}$ is flat
between zero and some large value where our quadratic approximation
breaks down.
Changing variables, the prior distribution for the dark matter ratio
$\d$ in the scenario with a large axion decay constant is
\be
{ dP \over d {\d} }\propto {1 \over \sqrt{\d}} \ \ \ {\rm (prior).}
\ee

\section{Anthropic Considerations}
We now compute the effect of increasing the dark matter abundance on
the number of observers. The analysis in this section is independent
of the details of the dark matter. However, inspired by axion
models, 
we hold fixed the
baryon to photon ratio, and simply increase the amount of dark
matter. This means that at decoupling, the density of baryons is
independent of the dark matter abundance, while the total matter
density at decoupling increases. Since we are assuming a flat
universe, the Hubble parameter at decoupling will
increase as the dark matter abundance increases.

We first compute the number of baryons inside the horizon as a
function of time, then discuss the number of observations per
baryon. 

\subsection{Number of baryons inside the horizon}
We are simply counting the number of baryons in the backward lightcone
of future infinity as a function of time.\footnote{Smaller causal diamonds have been used in
\cite{harnik}, but the fundamental definition of the causal diamond measure is to
count everything within a large causal diamond which extends beyond
our own bubble. This is equivalent, within our bubble, to counting
everything in the backward lightcone of future infinity.}
The number of baryons inside the horizon can be computed in the
homogeneous approximation. Assuming a flat universe, the metric is
\be
ds^2 = - dt^2 + a^2(t) (dx^2 + dy^2 + dz^2)
\ee
We begin at cosmological constant domination and work backwards.
 By
definition, $t = t_\Lambda$ when $\rho_m = \rho_\Lambda$. Therefore,
\be
\rho_b (t_\Lambda) = \rho_\Lambda / (1 + \d)~.
\ee
The horizon volume at the time $t_\Lambda$ is set by the de Sitter
radius, independent of the dark matter abundance. Therefore the baryon
mass inside the horizon at the time $t_\Lambda$ is
\be
M_b (t_\Lambda) \propto 1/(1 + \d)~.
\ee
More generally, since observations can be made at any time, we need
the number of baryons inside the horizon as a function of time. 
Note that both the Hubble parameter at vacuum domination and the
total matter energy density $\rho_m$ at vacuum domination are independent of the dark
matter ratio $\d$ in the approximation that radiation makes up a
negligible fraction of the energy density at \tl. 
Therefore, the scale factor as a function of time
from vacuum domination, $a(t - \tl)$, and the total matter density,
$\rho_m(t - \tl)$, are independent of the ratio \d. Since the scale
factor is unaffected by \d, the volume inside the horizon, $V(t -
\tl)$ is also independent of \d. On the other hand,
the energy density in radiation, $\rho_\gamma$, does depend on \d. 

Therefore, as long as radiation is a negligible fraction of the energy
density, the total mass inside the horizon, $M(t - \tl)$, is {\it
  independent} of the dark matter ratio \d. So the number of baryons
inside the horizon, for all times with negligible energy density in
radiation, is given by
\be
N_b (t- \tl) = {1 \over 1 + \d} N_0 (t - \tl)
\ee
where $N_0 (t - \tl)$ is a universal function which does not depend on
\d.

Of course the dark matter ratio does affect the early universe. The
baryon to photon ratio is fixed, so at decoupling the ratio $\rho_b/
\rho_\gamma$ 
is independent of $\d$. So the total
matter density at decoupling depends on \d, 
\be
\rho_m (t_{DC}) \propto 1 + \d
\ee
Therefore matter-radiation equality happens at a higher temperature as
\d\ is increased. Quantitatively, the matter-radiation ratio redshifts
as
\be
{\rho_m \over \rho_\gamma} \sim a
\ee
and the temperature redshifts as $T \sim 1/a$,
so
\be
T_{eq} \propto 1 + \d ~.
\ee
The matter density is also higher at equality,
\be
\rho_m (t_{eq}) \propto T_{eq}^4 \propto (1 + \d)^4
\ee

To summarize, the bottom line is that \d\ affects the early universe, changing the
time and temperature at matter-radiation equality. But for all values
of \d\ the universe matches onto a universal late time behavior (in
the homogeneous approximation) as
soon as the energy density in radiation is much less than than the
energy density in matter. During this era, $a(t- \tl)$ and $\rho_m (t
- \tl)$ are both universal functions which are independent of \d, so
the total mass inside the horizon $M(t - \tl)$ is also independent of
\d. So the number of baryons at any time with negligible energy
density in radiation is a universal function times the factor $1/(1 + \d)$.


\subsection{The number of observations per baryon}
Now in principle, we should count the number of observations inside
the causal patch by
multiplying the number of baryons by the number of observations per
baryon per unit time:
\be
N_{obs} = \int dt N_b (t) f(t)
\ee
where $f(t)$ is the number of observations per baryon per unit
time. One estimate for the function $f(t)$ is that it is
proportional to the collapsed baryon fraction. We will not try to
compute $f$ in detail here, because it is quite difficult and much
work has already been done. 

To proceed, we make a crude approximation. First we need a lower bound
on the amount of dark matter. As the amount
of dark matter is decreased, the size of the biggest nonlinear
structures decreases. According to Tegmark, Aguirre, Rees, and Wilczek
\cite{tarw}, density perturbations at scales similar to our galaxy
become nonlinear only if
\be
Q \gtrsim 10^{-5} \left( 1 + \d_0 \over 1 + \d \right)^{4/3}
\ee
where $\d_0 \approx 5$ is the dark matter ratio in our universe and
$Q \approx 2 \times 10^{-5}$ is the magnitude of the density perturbations. This
criterion gives
\be
\d_{\rm min} \approx 2.5 ~.
\ee
It is interesting to note that while the above bound depends on $Q$ and $\Lambda$,
\cite{tarw} give other bounds on the dark matter ratio which are more
general. 
For example, struture formation is seriously
impeded by Silk damping when $\d \lesssim 1$, independent of both $Q$
and $\Lambda$, so a lower bound in
the neighborhood of $\d \approx 1$ is somewhat robust.

What is the upper bound at which the amount of dark matter starts to
seriously impair observer formation? This is a controversial subject,
and in any case the most stringent upper bound in the literature is
that of \cite{tarw} who find $\d \lesssim 10^2$. On the other hand,
Hellerman and Walcher \cite{simeon} find no compelling evidence for a bound stronger
than $\d \lesssim 10^5$. One additional consideration in the causal
diamond measure is that the number of baryons inside the causal
diamond becomes very small after the cosmological constant dominates
the vacuum energy, so if for large \d\ observer formation is delayed
past the time \tl, the probability of observing such large \d\ will be suppressed.

Here, we will use the results of \cite{tarw} and approximate the
number of observers per baryon as a constant in the range
\be
2.5 < \d < 100
\ee
while assuming that there are approximately zero observers per baryon
outside this range. We will also
demonstrate that the prediction is robust against changing the
assumptions.
Clearly, there is room for improvement in these considerations.

\section{Prediction for the Dark Matter Abundance}
Now we can combine the known prior distribution for axion dark matter
with the anthropic counting of observers to generate the probability
distribution for the observed dark matter abundance. To get the final
probability distribution for \d, we multiply the prior
distribution
\be
{d P \over d \d} \propto {1 \over \sqrt{\d}} \ \ \ {\rm (prior)}
\ee
by the number of baryons inside the causal patch,
\be
N_b \propto {1 \over 1 + \d}
\ee
and the number of observations per baryon,
\bea
{N_{obs} \over N_b} &\approx& {\rm constant,\ for}\ \ 2.5 < \d < 100
\nonumber \\
{N_{obs} \over N_b} &\approx& 0 \ \ \ \ \ \ \ \ \ \ \ \ \ \ \ \ \ \ {\rm otherwise.}
\eea
Thus the final probability distribution
for the dark matter abundance
\bea
{d P \over d \d} &\propto&
 {1 \over \sqrt{\d} (1 + \d)} \ \ \ \ 2.5 < \d < 100 \nonumber \\
{d P \over d \d} &\approx& 0 \ \ \ \ \ \ \ \ \ \ \ \ \ \ \ \ \ {\rm otherwise}
\label{dist}
\eea
This probability distribution is pictured in figure 1, along with the
``conventional'' anthropic probability distribution, which instead
counts the number of observers per baryon.
\begin{figure}[!htb]
\center
\includegraphics [scale=1]
{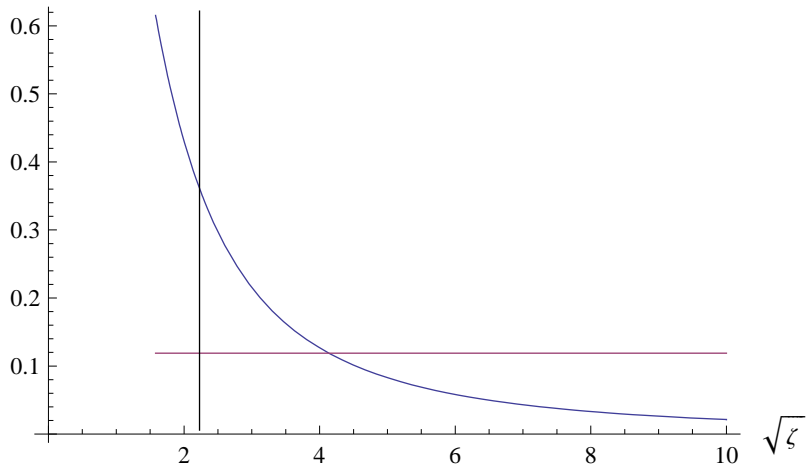} \caption{The probability distribution for the dark matter ratio \d\
is plotted as a function of $\sqrt{\d}$. The conventional anthropic analysis
gives approximately a flat distribution in the range $2.5 \lesssim \d
\lesssim 100$ (horizontal line), while the causal diamond measure
gives an additional factor of $1/(1 + \d)$ (curved line). The vertical
line shows the observed value $\d = 5$. In the conventional analysis
only about 8\% of observers form in regions with less dark matter than
we observe, even with the most favorable assumptions about observer
formation.  In
the causal diamond measure the observed value is fairly typical:
30\% of observers form in regions with less dark matter than we observe. 
The causal diamond
prediction is rather insensitive to assumptions about observer
formation at large \d.}
\label{fig-pot}
\end{figure}

To address the typicality of our observations, it makes sense to
compute how much of the probability distribution is on each side of
the observed value. Integrating (\ref{dist}), we find that 30\% of
the probability is at smaller \d, while 70\% is at larger \d, so our
observations are quite typical. The probability distribution is fairly
broad: 68\% of observers form in regions with $\d \lesssim 15$, and
95\% of observers form in regions with $\d \lesssim 65$.

Since observer formation at large \d\
is poorly understood, let's see what happens to our distribution if we
asssume that the number of observations per baryon is constant all the
way up to $\d = \infty$. With this assumption, 25\% of observers form
in regions with less dark matter than us, so our observations are
still fairly typical. Since the prediction is not significantly
affected, it is not urgent to understand observer formation for very
large dark matter ratios $\d > 100$.

To probe the sensitivity of our results further, assume for a moment
that the number of observers per baryon is constant for $1 < \d <
100$. With this assumption, 53\% of observers form in regions with
less dark matter than we have. While it is probably not realistic to
think that structure formation in our universe can proceed effectively
down to $\d \approx 1$, as we mentioned in the previous section once
$Q$ and $\Lambda$ are allowed to vary the lower bound $\d_{min}
\approx 2.5$ is not valid in general, while the lower bound $\d_{min}
\approx 1$ is valid \cite{tarw}.

\section{Future Directions}
It will be interesting to extend these considerations to a larger
landscape in which more parameters are allowed to vary. It is
particularly interesting to ask what happens once the cosmological
constant and the density contrast $Q$ are allowed to vary along with
the dark matter ratio.
A more precise prediction within the small landscape studied here would
require an improved analysis of observer formation for a range of dark
matter ratios $0 < \d \lesssim 1000$, although as we explained large
values of \d\ have only a small effect on the prediction.

In supersymmetric field theory and string theory models, one has to
worry about cosmological problems arising from the Saxion and other
moduli \cite{saxion}. It is important to verify that the positive
results found here survive in a more detailed model.  More generally,
it would be interesting to repeat the statistical arguments here in
the presence of other sources of dark matter in addition to the axion.

An additional interesting direction is the search for observational
confirmation of these ideas. As discussed recently by \cite{thomas,
  htw, kapnel}
axionic dark matter
has observational signatures in the form of isocurvature
perturbations; these signals are enhanced when the axion misalignment
angle is ``unnaturally'' small due to anthropic selection \cite{kapnel}. 
These perturbations can be produced during inflation if
the inflation scale is high enough. They can also be produced prior to
inflation and survive observationally if inflation does not go on for
too long.  In fact, Kaplan and Nelson \cite{kapnel} conclude that for
some parameter choices, isocurvature perturbations from the
preinflationary era will be observable if there are fewer than about 74
efoldings of slow roll inflation. (This is in conventions where the
bound coming from the observed flatness of the universe corresponds to
60 efoldings.) Combined with arguments \cite{fkms}
that inflation is quite likely to last for fewer than 74 efoldings, 
this is a very exciting
possibility which deserves further study.
				
\subsubsection*{Acknowledgements}
I would like to thank Raphael Bousso, Simeon Hellerman, Ann Nelson, 
Stefan Leichenauer, Leonard Susskind, and especially Jens Niemeyer for
helpful discussions.  This work was supported by the Berkeley Center
for Theoretical Physics, by NSF CAREER grant 0349351,
and by DOE grant DE-AC02-05CH11231.

\end{document}